\documentclass{PoS}

\pdfoutput=1

\usepackage{graphicx}
\usepackage{subfigure}
\usepackage{url}

\newcommand{\bb}{$b\bar{b}$}
\newcommand{\ppbar}{$p\bar{p}$}
\newcommand{\Hbb}{$H \rightarrow b\bar{b}$}

\newcommand{\ggH}{$gg \rightarrow H$}
\newcommand{\invfb}{\mbox{fb$^{-1}$}}

\newcommand{\WHlvbb}{$WH \rightarrow \ell \nu b\bar{b}$}
\newcommand{\ZHllbb}{$ZH \rightarrow \ell \ell  b \bar{b}$}
\newcommand{\ZHvvbb}{$ZH \rightarrow \nu \nu  b \bar{b}$}
\newcommand{\mH}{$m_{H}$}

\newcommand{\gevcc}{GeV/$c^2$}

\def\Hgg{$H~\rightarrow \gamma \gamma~$}

\newcommand{\HWW}{$H \rightarrow WW$}
\newcommand{\HWWlvlv}{$H \rightarrow WW \rightarrow \ell \nu \ell \nu$}

\newcommand{\Et}{\mbox{$E_T$}}

\newcommand{\met}{\mbox{$\protect \raisebox{0.3ex}{$\not$}\Et$}}

\title{Combination of Standard Model Higgs searches at CDF}

\ShortTitle{Combination of Standard Model Higgs searches at CDF}

\author{\speaker{Karolos Potamianos}%
       \thanks{On behalf of the CDF Collaboration}\\
       Purdue University\\
       Fermilab, Batavia, IL, USA\\
       E-mail: \email{karolos@fnal.gov}}

\abstract{
We present the latest combination of searches for a standard model (SM) Higgs boson in \ppbar\ collisions at $\sqrt{s}= 1.96$ TeV recorded by the CDF~II detector at the Fermilab Tevatron. Using data corresponding to 2.3-5.9 \invfb\ of integrated luminosity, we perform searches in a number of different production and decay modes and then combine them to improve sensitivity. No excess in data above that expected from backgrounds is observed; therefore, we set upper limits on the production cross section times branching fraction as a function of the SM Higgs boson mass (\mH). The combined observed (expected) limit is 1.9 (1.8) times the SM prediction at $ m_H =  115$~\gevcc\ and 1.0 (1.1) times the SM prediction at $m_H = 165$~\gevcc.}

\FullConference{35th International Conference of High Energy Physics\\
		 July 22-28, 2010\\
		 Paris, France}

\begin{document}

\section{Introduction}

\begin{figure}
\centering
\subfigure[]{\includegraphics[width=.28\linewidth,height=.24\linewidth]{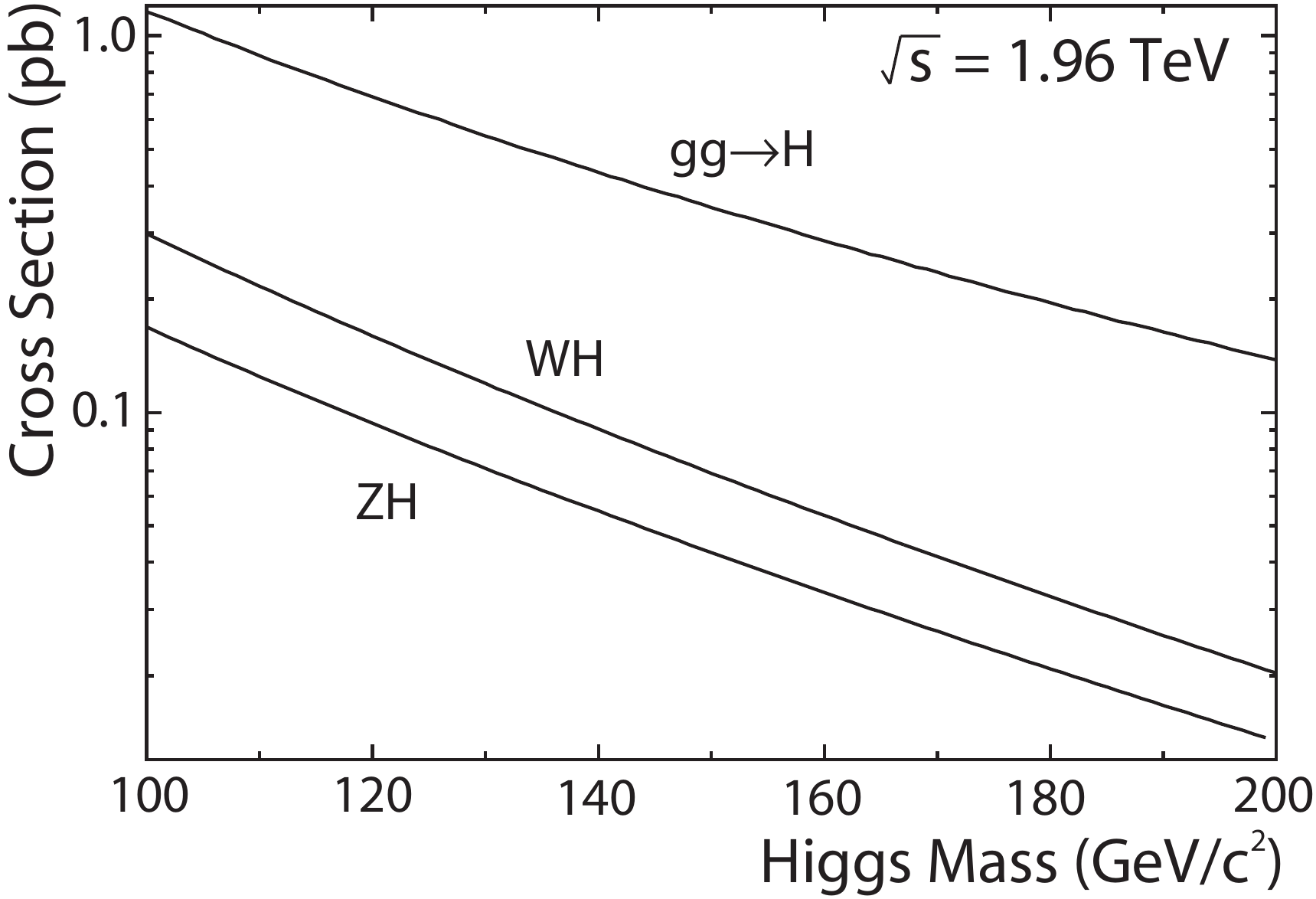}}
\subfigure[]{\includegraphics[width=.28\linewidth,height=.28\linewidth]{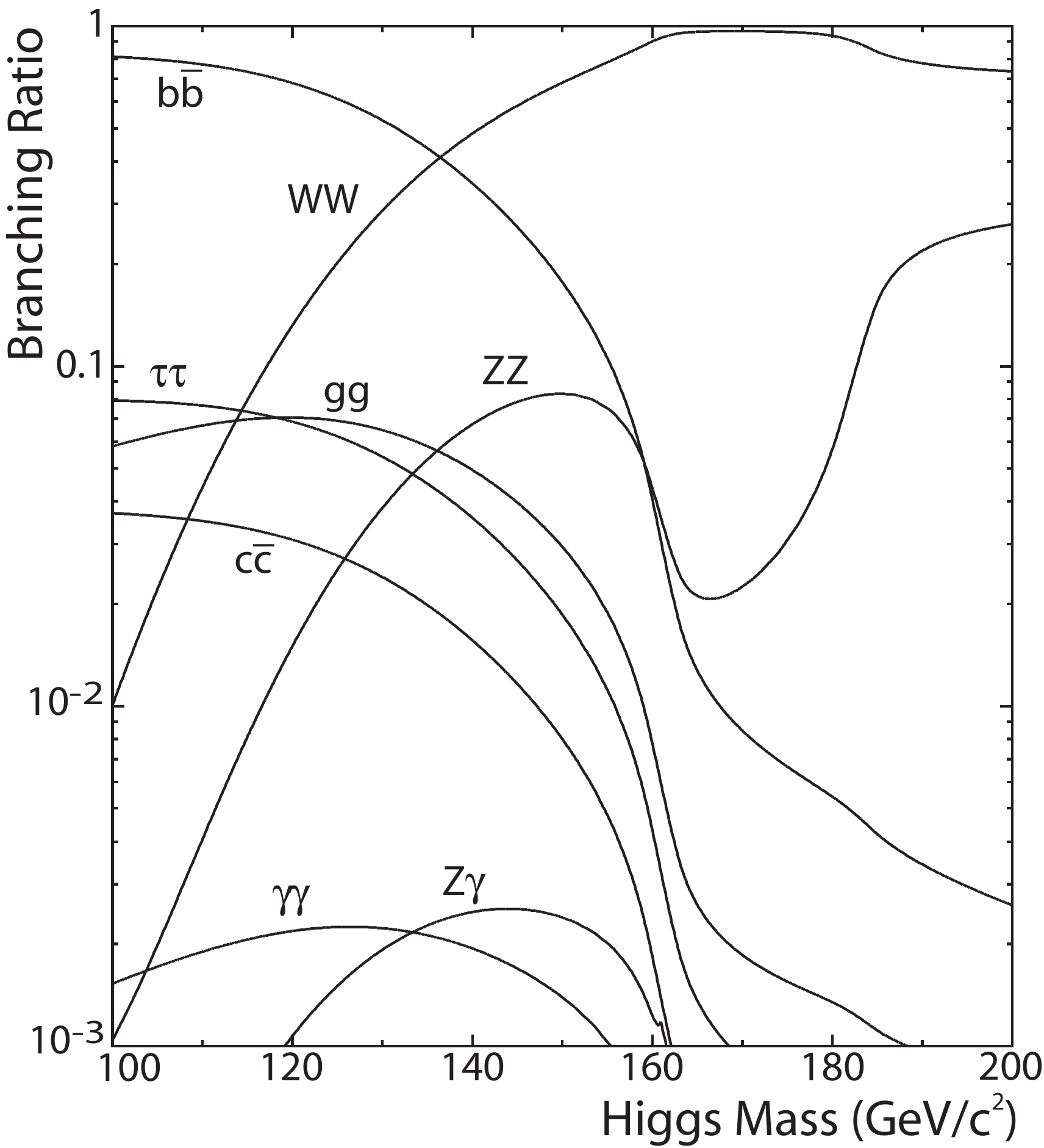}}
\caption{Main SM Higgs production cross-sections (a) and branching ratios (b) in \ppbar\ collisions at the Tevatron ($\sqrt{s} = 1.96$~TeV).}
\label{fig:SMHiggsXSecAndDecay}
\end{figure}

Understanding the mechanism of electroweak symmetry breaking has been a major goal of the high energy physics community for several decades. 
The BEHHGK mechanism~\cite{ref:BEHHGK} proposed in 1964 added the Higgs boson to the standard model (SM) of particle physics. This new particle has yet to be observed. 
Direct searches at the LEP experiments have placed a lower limit of $114.4$~\gevcc\ on the SM Higgs boson mass (\mH) at 95\% confidence level (C.L.), while precision electroweak measurements place an indirect limit of $ m_H < 158$~\gevcc\ at 95\% C.L.~\cite{ref:LEPEWWG}
Figure~\ref{fig:SMHiggsXSecAndDecay} shows the main production and decay modes of the SM Higgs boson (H) at the Tevatron, a \ppbar\ collider at $\sqrt{s} = 1.96$~TeV located at Fermilab.
The recent observation of single top~\cite{ref:single_top} and diboson~\cite{ref:diboson} production in semi-leptonic decays have prepared the way for Tevatron to probe processes with  sub-picobarn cross sections, among which the SM Higgs boson has the main focus. This is now a central part of the Tevatron program.
We present the status of the combination of multiple direct searches for the SM Higgs boson using up to $5.9$~\invfb\ of data collected by the CDF~II detector~\cite{ref:CDF2}.

\section{Standard Model Higgs searches at CDF}

The search for the SM Higgs boson is challenging due to the small signal expectation and the large backgrounds\footnote{At the Tevatron, dijet QCD events are produced at a rate ten orders of magnitude higher than the SM Higgs boson.}. The CDF search strategy is based on a {\it divide and conquer} approach: since no single signature has sufficient sensitivity, we divide the search into many different channels. Each channel exploits a decay mode using dedicated triggers and analysis techniques. There are 56 such channels in total. 
According to Figure~\ref{fig:SMHiggsXSecAndDecay}, \Hbb\ is the dominant mode for $ m_H < 135$~\gevcc\ (low mass) while \HWW\ dominates at high mass. Because of the overwhelming irreducible QCD background, it is necessary to seek a striking signature to identify the SM Higgs boson. At low mass, we focus on processes where it is produced in association with a $W$ or $Z$ boson, while at high mass, the decay products of the $W$ allow probing of the five times more likely direct production \ggH.

At low mass, the main channels are dedicated to events with zero (\met+\bb), one ($\ell\nu$\bb) or two ($\ell\ell$\bb) identified leptons from the following decay modes: \ZHvvbb, \WHlvbb\ and \ZHllbb. At high mass, the \HWWlvlv\  decay path reduces background and improves signal purity. Other channels include $VH\rightarrow jj$\bb, an inclusive $H\rightarrow \tau \tau$, and a new \Hgg\ analysis. These searches are described in detail in other proceedings to this conference~\cite{ref:OtherContributions} and online~\cite{ref:ResultsHiggsCDF}. All these results are combined to determine the sensitivity of the CDF experiment to the SM Higgs boson.

Each individual analysis makes use of multivariate techniques to exploit the information in each event. It is crucial to check that all the inputs to these techniques as well as their outputs are well described by the background models. We perform these checks comparing the background model to the data in various control regions. Each of these regions is defined to check the modeling of a major background component. The CDF Collaboration has built confidence over the years in detector modeling and multivariate techniques have been successfully used in recent observations~\cite{ref:single_top,ref:diboson}.

\section{Combination of Standard Model Higgs searches at CDF}

We combine the results from each individual analysis using a Bayesian approach. Like for each individual channel, we rely on the distribution of final discriminants, and not just on event counts. We use a binned likelihood calculation based on Poisson probabilities to extract an upper limit on the production of the SM Higgs boson. Systematic uncertainties are treated as nuisance parameters with truncated Gaussians. Two types of systematic uncertainties are considered, which affect the {\it rate} and {\it shape} of the estimated signal and background distributions in a correlated way.  Sources of systematic uncertainties include the luminosity, the theoretical cross sections, scale and PDF variations, lepton identification, $b$-jet identification, jet energy scale (JES) and detector effects.

\section{Results and Future Prospects}

\begin{figure}
\centering
\subfigure[]{\includegraphics[width=0.48\textwidth]{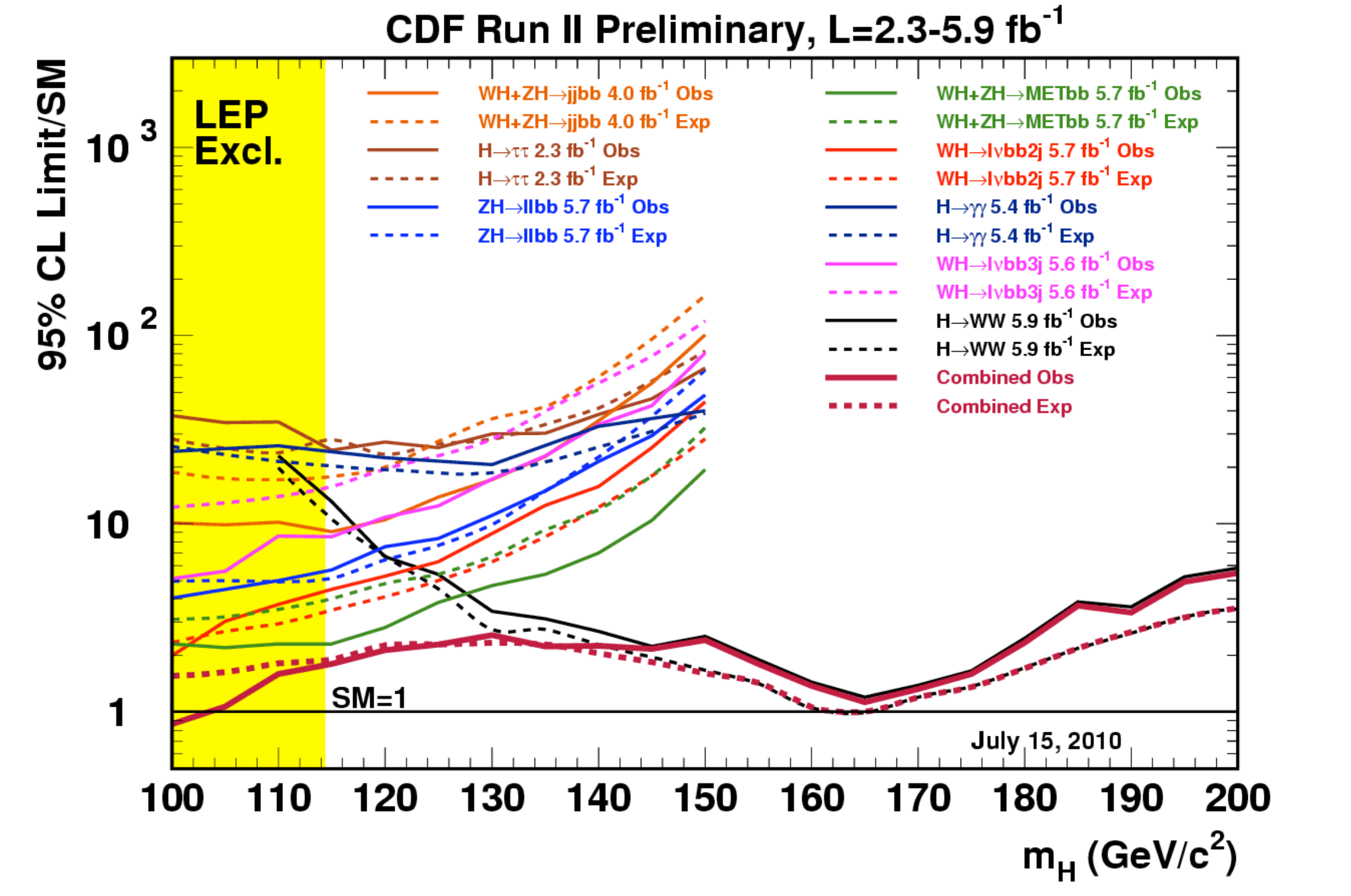}}\\ \vspace{-.2in}
\subfigure[]{\includegraphics[width=0.44\textwidth]{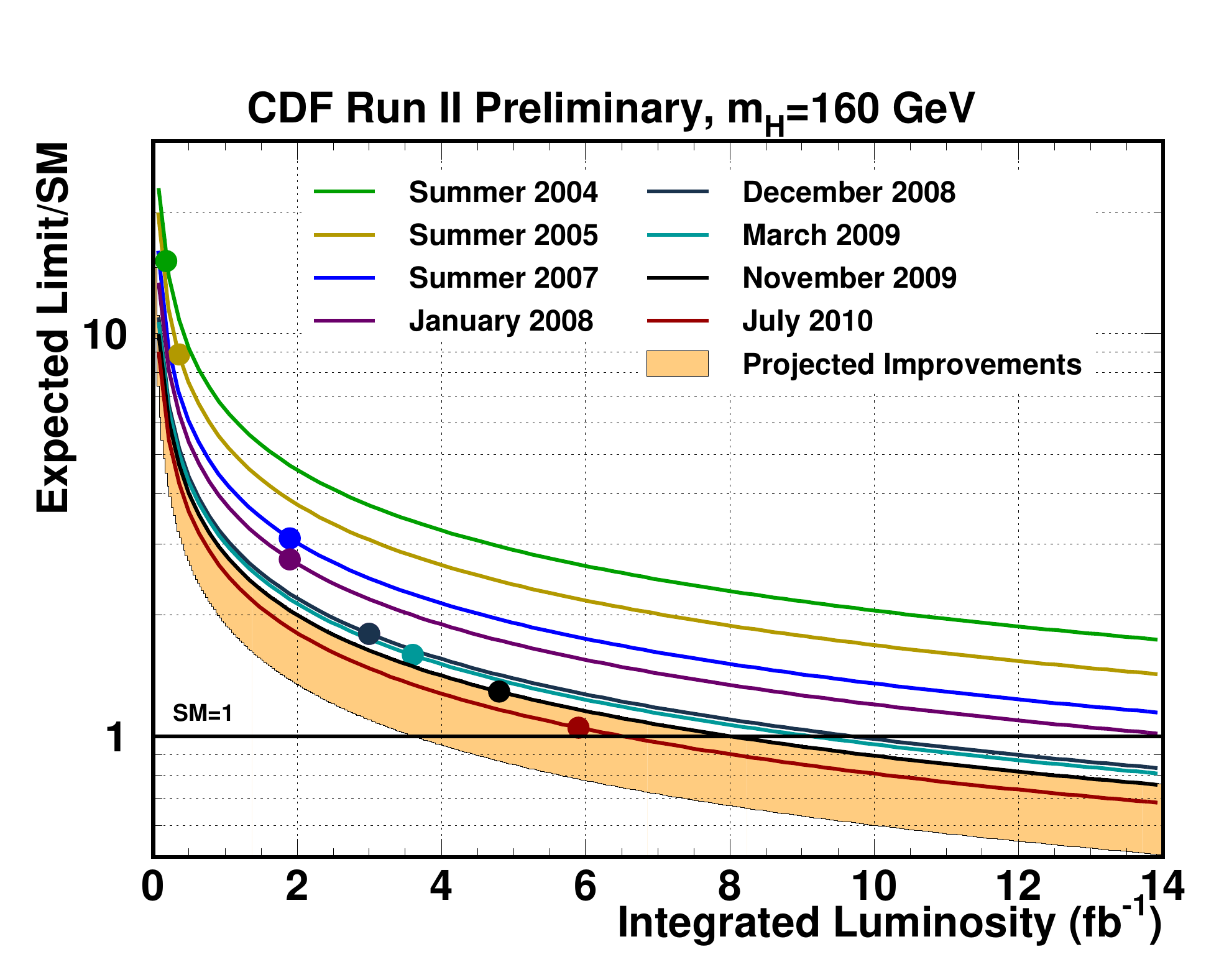}}
\subfigure[]{\includegraphics[width=0.48\textwidth]{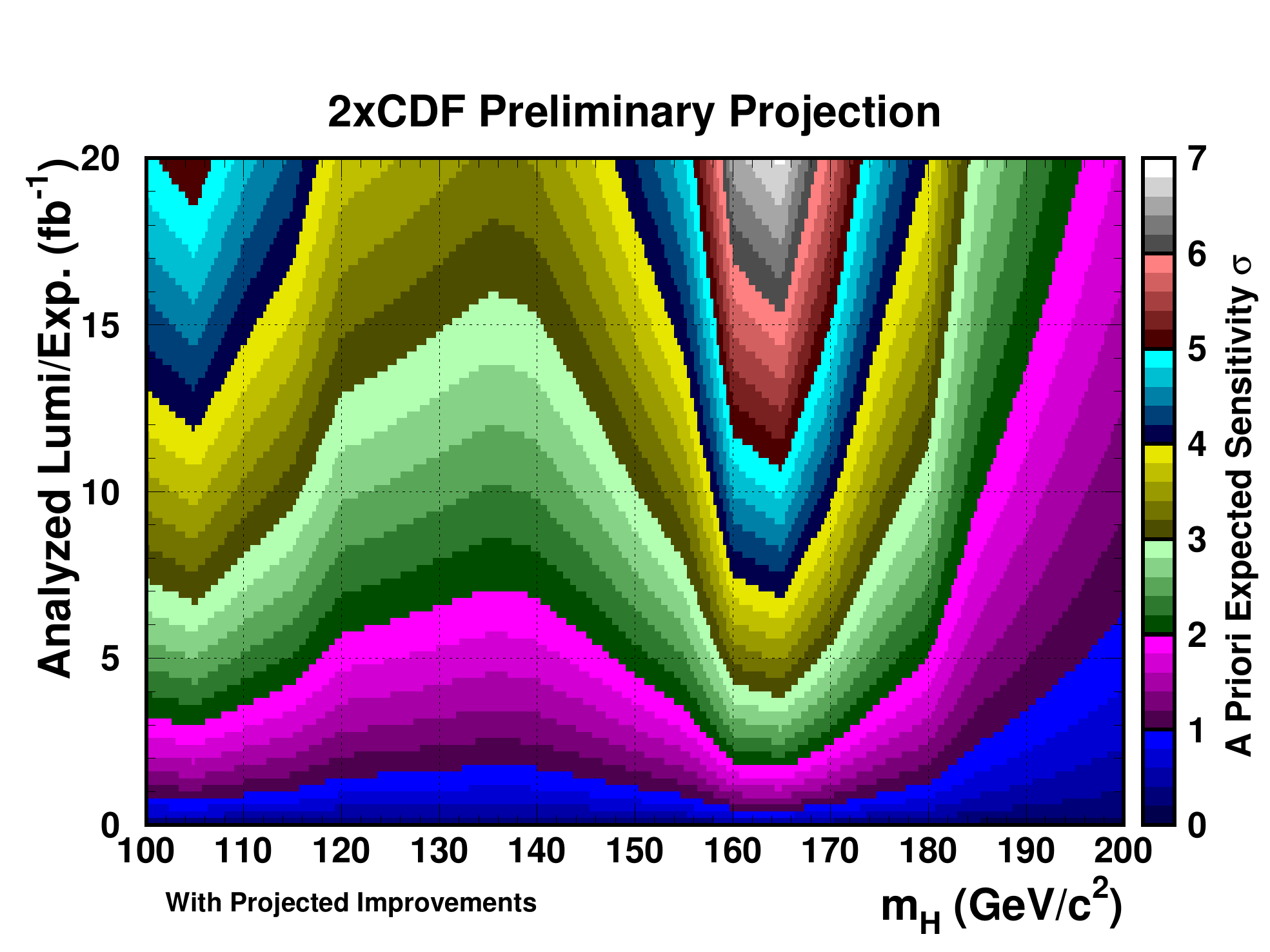}}
\caption[]{ (a) Limits at 95\% C.L. on the SM Higgs boson production cross section times the SM prediction for the individual channels and the CDF combined result. (b) Analysis improvements in the search for the SM Higgs boson. The shaded band is based on two stages of 50\% improvement of the summer 2007 result. The July 2010 result (red) is within expectations. (c) Expected sensitivity to the SM Higgs boson assuming both Tevatron experiments perform as CDF, and not including information from CDF data. }
\label{fig:CDF_improvements}
\end{figure}

The observed and expected limits on the SM Higgs boson cross section are shown in units of the SM prediction and as a function of \mH\ for each individual channel in Figure~\ref{fig:CDF_improvements}(a). We combine these results and find expected limits ranging from 1 to about 4 times the prediction of the SM rate, with all limits below 2.3 for $ m_H < 185$~\gevcc. The combined observed (expected) limit is 1.9 (1.8) times the SM prediction at \mH\ $=  115$~\gevcc\ and 1.0 (1.1) times the SM prediction at $m_H= 165$~\gevcc.

A combination with the D\O\ experiment excludes a SM Higgs boson with \mH\ between $158$ and $175$~\gevcc\, and between $100$ and $109$~\gevcc\ at 95\% C.L. Detailed information on this combination can be found in another contribution to these proceedings~\cite{ref:TevComb}.

With more data to analyze and improvements still being added to the analyses (Figure~\ref{fig:CDF_improvements}(b)), the Higgs boson search results will continue to be an exciting topic until the end of the Tevatron run.  With a full dataset expected to be on the order of 10~\invfb, there is a significant chance (Figure~\ref{fig:CDF_improvements}(c)) that the Tevatron will see some evidence of the elusive SM Higgs boson.   

\end{document}